\documentclass[12pt]{article}

\makeatletter
\def\eqnarray{%
 \stepcounter{equation}%
 \let\@currentlabel=\theequation
 \global\@eqnswtrue
 \global\@eqcnt\z@
 \tabskip\@centering
 \let\\=\@eqncr
 $$\halign to \displaywidth\bgroup\@eqnsel\hskip\@centering
 $\displaystyle\tabskip\z@{##}$&\global\@eqcnt\@ne
 \hfil$\displaystyle{{}##{}}$\hfil
 &\global\@eqcnt\tw@$\displaystyle\tabskip\z@{##}$\hfil
 \tabskip\@centering&\llap{##}\tabskip\z@\cr}
\makeatother

\begin{document}

\title{\bf More on the Triplet Killing Potentials of \\
Quaternionic K\"ahler manifolds}

\author{Shogo Aoyama\thanks{e-mail: spsaoya@ipc.shizuoka.ac.jp} \\
       Department of Physics \\
              Shizuoka University \\
                Ohya 836, Shizuoka  \\
                 Japan}
                 
\maketitle

\vspace{2cm}
                 
\begin{abstract}
We show the properties of the triplet Killing potentials of quaternionic K\"ahler manifolds which have been missing in the literature. It is done by means of the metric formula of the manifolds. 
 We compute the triplet Killing potentials  for the quaternionic K\"ahler manifold $Sp(n+1)/Sp(n)\otimes Sp(1)$ as an illustration.

\end{abstract}

\vspace{5cm}



\newpage

It is well-known that when hypermultiplets couple with ${\cal N}$=$2$ supergravity in four dimensions, the moduli space  is  required to be  a quaternionic  K\"ahler manifold\cite{BW,BLP}.
A quaternionic K\"ahler manifold is a $4n$-dimensional real manifold. 
The holonomy group of quaternionic K\"ahler manifolds is $Sp(1)\otimes Sp(n)$ and the $Sp(1)$ conection must not vanish. 
 When the gravitational coupling $\kappa \rightarrow 0$, this $Sp(1)$ conection  is required to vanish, the holonomy group gets contained in $Sp(n)$ and 
the quaternionic K\"ahler manifold approaches to a hyper K\"ahler manifold\cite{AF}.
 Such a quaternionic or  hyper K\"ahler manifold appears as a moduli  space for type II superstring compactification on Calabi-Yau manifolds, together with  a special K\"ahler manifold given by vector multiplets\cite{BLP}.  The direct product of both moduli spaces  is an arena for the mirror symmetry in the type II superstring\cite{Sei}. 

Quaternionic K\"ahler manifolds are by now well-studied subjects\cite{BW}\cite{W}-\cite{energy}. The existence of the triplet Killing potentials $\vec M^A$ is a hallmark of quaternionic K\"ahler manifolds when they are group manifolds. The properties of the $\vec M^A$ were well studied in the literature\cite{Killing}-\cite{energy}. But there are still important properties which are missing in generality of the arguments. 
In this letter we show them, explicitly constructing the metric of quaternionic K\"ahler manifolds. With the specific metric at hand the geometric quantities of the manifolds like the metric, the Riemann curvature {\it etc}, are given in terms of  the triplet Killing potentials alone. 
Among them  we stress on the relations implying that 
 $\vec M^A$ is an eigen vector of the Beltrami-Laplace operator for the manifold and the quantity $\vec M^A\cdot \vec M^A$, which was called {\it energy} in \cite{energy}, is a constant related with  the Riemann scalar curvature.  This is justified  as long as the metric of quarternionic K\"ahler group manifolds manifolds is non-degenerate. Similar relations  were known for the  Killing potentials which  exist for the ordinary K\"ahler group manifolds\cite{BW2}-\cite{AM}. 
We develope our arguments noting a  pararellism between both K\"ahler group manifolds. Finally we  compute the triplet Killing potentials for the quarternionic K\"ahler coset space $Sp(n+1)$ $/Sp(n)\otimes Sp(1)$ to illustrate our general arguments.

We start with a summary of the geometric properties discussed in \cite{Killing}-\cite{energy}. 
A quaternionic K\"ahler manifold is endowed with a triplet complex structure $\vec J_a^{\  b}$. Assume that it is a group manifold admitting an isometry $G$ realized by Killing vectors $R^{Aa}$ with $A=1,2,\cdots,{\rm dim}\ G$. For the quantity  $\vec J_a^{\  b}R^A_{\ b}(\equiv \vec J_a^{\  b}g_{bc}R^{Ac})$ we have   triplet Killing vectors $\vec M^A$ such that 
\begin{eqnarray}
\nabla_a \vec M^A =\nu \vec J_a^{\  b}R^A_{\ b}, \quad\quad
 \nu \equiv {R\over 4n(n+2)}, \label{dM}
\end{eqnarray}
with the Riemann scalar curvature $R$. 
It is given by 
\begin{eqnarray}
 \vec M^A = \vec
 r^A + R^{Aa}\vec \omega_a, \label{MK}
\end{eqnarray}
with a spin connection $\vec\omega_a$ and an appropriate $\vec r^A$ which will be explained soon later.  $\vec M^A$ can be also written in the form 
\begin{eqnarray}
 \vec M^A = {1\over 2n}\vec J_a^{\  b}\nabla_b R^{Aa} \label{dR}
\end{eqnarray}
and satisfies the relation
\begin{eqnarray}
\nu R^A_{\ a}R^B_{\ b}\vec J^{ab} = f^{ABC}\vec M^C + \vec M^A\times \vec M^B. 
\label{4}
\end{eqnarray}
Here  $f^{ABC}$ are the structure constants of $G$ 
and $\vec J^{bc}\equiv g^{bd}\vec J_d^{\ c}$.

We remember quite similar arguments for  K\"ahler manifolds\cite{BW}\cite{A1}-\cite{AM}. Namely a K\"ahler manifold is endowed with a singlet complex structure $J_a^{\ b}$. 
It may be locally set to be 
\begin{eqnarray}
J_a^{\ b} = 
\left(
\begin{array}{cc}
 -i\delta_\alpha^{\ \beta}  &  0 \\
           &    \\
         0      &  i\delta_{\overline \alpha}^{\ \overline \beta} \\
\end{array}\right).   \nonumber
\end{eqnarray}
If the K\"ahler manifold admits an isometry $G$, realized by (anti-)holomorphic Killing vectors $R^{A\alpha}(\bar R^{A\bar\alpha})$, there exist Killing potentials such that 
\begin{eqnarray}
\partial_\alpha M^A = -ig_{\alpha\bar \beta}\bar R^{A\bar\beta}, \quad \quad
\partial_{\bar\alpha}M^A = ig_{\beta\bar\alpha}R^{A\beta}. \label{dM'}
\end{eqnarray}
It was shown that such Killing potentials are given by
\begin{eqnarray}
 M^A &=& -{i \over {\cal N}_{adj}}\nabla_\alpha R^{A\alpha} \Bigl(={i \over {\cal N}_{adj}}\nabla_{\bar\alpha}\bar R^{A\bar\alpha}\Bigr),
 \label{dR'}  \\
or \hspace{2cm} M^A &=& i( R^{A\alpha}\partial_\alpha K -F^A), \hspace{3cm}
  \label{MK'}  
\end{eqnarray}
and satisfy the relation 
\begin{eqnarray}
  -iR^{B\beta} \bar R^{C\bar\gamma}g_{\beta\bar\gamma} 
  +iR^{C\beta} \bar R^{B\bar\gamma}g_{\beta\bar\gamma}
= f^{ABC} M^A.  \label{fM'} 
\end{eqnarray}
In (\ref{dR'})  we have used the normalization  $f^{ABC}f^{ABD}=2{\cal N}_{adj}\delta^{CD}$. 
In (\ref{MK'}) $K$ is the K\"ahler potential and $F^A(\bar F^A)$ are (anti-)holomorphic quantities  that one may find from the Lie-variation 
\begin{eqnarray}
{\cal L}_{R^A} K = F^A + \bar F^A.
\end{eqnarray}

Eqs (\ref{dM})$\sim$(\ref{4}) of  quaternionic K\"ahler group manifolds are so similar to (\ref{dM'})$\sim$(\ref{fM'}) of  K\"ahler group manifolds.  About $M^A$ of  K\"ahler group manifolds we know more properties than those. By the Lie-variation we have\cite{BW} 
\begin{eqnarray}
{\cal L}_{R^A} M^B = f^{ABC}M^C. \label{LieM}
\end{eqnarray}
From (\ref{dM'}) and (\ref{dR'}) it follows that $ M^A$ is an eigen vector of the Beltrami-Laplace operator as\cite{AM}
\begin{eqnarray}
g^{\alpha\bar \beta}\nabla_\alpha\nabla_{\bar\beta}M^A= -{\cal N}_{adj}M^A \label{BLdif}
\end{eqnarray}
Moreover, when  K\"ahler manifolds are irreducible coset spaces, then 
  the geometric quantities can be written in terms of the Killing potentials\cite{A1,A2} 
\begin{eqnarray}
g_{\alpha\bar\beta}&=& \partial_\alpha M^A\partial_{\bar\beta}M^A,  \label{gMM}   \\ 
 R_{\alpha\bar\beta\gamma\bar\delta}&=& -\partial_\alpha\partial_{\bar\beta}M^A\partial_\gamma\partial_{\bar\delta}M^A   
 =f^{ABE}f^{CDE}R^A_\alpha R^B_{\bar\beta}R^C_\gamma R^D_{\bar\delta},  
\label{RiemannM}  \\ 
 R&=& -R_{\alpha\bar\beta\gamma\bar\delta}g^{\alpha\bar\beta}g^{\gamma\bar\delta} = {\cal N}_{adj}^2 M^AM^A.    \label{ScalarRiemann}
\end{eqnarray}
The geometrical properties of  quaternionic K\"ahler manifolds which correspond to (\ref{LieM})$\sim$ (\ref{ScalarRiemann}) are missing in the literature\cite{Killing}-\cite{energy}. 
In this letter we show them by completing   the  pararellism between both K\"ahler group manifolds. They are 
\begin{eqnarray}
 {\cal L}_{R^A}\vec M^B &=& f^{ABC}\vec M^C + \vec r^A\times \vec M^B, 
\label{LieM'}   \\
\nabla_a\nabla^a \vec M^A &=& -2n\nu \vec M^A, \label{BLd'}  \\
g_{ab}&=& {1\over 3\nu^2}\nabla_a \vec M^A\cdot\nabla_b\vec M^A, 
 \label{gMM'}  \\
R&=& {2\over 3}n(n+2)\vec M^A\cdot\vec M^A,    \label{ScalarRiemann'} \\
R_{abcd}&=& {1\over 3\nu^2}\Bigl[[\nabla_a,\nabla_b], \nabla_c\Bigr]\vec M^A\cdot\nabla_d\vec M^A   \nonumber \\
 &=&f^{ABE}R^A_{\ a} R^B_{\ b}f^{CDE}R^C_{\ c}R^D_{\ d}  \nonumber \\
 &\quad& 
- {1\over 2}f^{ABC}R^A_{\ a}R^B_{\ b}R^{Ce}f^{DEF}R^D_{\ c}R^E_{\ d}R^F_{\ e}  
\label{RiemannCurv'} \\
&\quad& -{1\over 4}f^{ABC}R^A_{\ d}R^B_{\ b}R^{Ce}f^{DEF}R^D_{\ c}R^E_{\ a}R^F_{\ e}    \nonumber \\
 &\quad&  + {1\over 4}f^{ABC}R^A_{\ d}R^B_{\ a}R^{Ce}f^{DEF}R^D_{\ c}R^E_{\ b}R^F_{\ e}. \nonumber
\end{eqnarray}
Here $\vec r^A$ is the same quantity that appeared in (\ref{MK}). 
These formulae  are useful when    four-fermi couplings and scalar potentials in ${\cal N}$=$2$ supergravity  are analyzed from a phenomelogical point of view, by identifying the isometry group $G$ or some subgroup  with a grand unification gauge group. 

We give a short review on quaternionic K\"ahler manifolds. Consider a real $4n$-dimension\-al Riemann manifold ${\cal M}$ with local coordinates $\phi^a =(\phi^1,\phi^2,\cdots,\phi^{4n})$. The line element of the manifold is given by 
$
ds^2 = g_{ab}d\phi^a d\phi^b.  
$
If ${\cal M}$ is a quaternionic manifold, there exists a triplet complex structure $\vec J_a^{\ b}\equiv ( J^{1\  b}_{\ a},J^{2 \ b}_{\ a},J^{3\  b}_{\ a})$ satisfying the property
\begin{eqnarray}
J^{\alpha\ b}_{\ a}J^{\beta\ c}_{\ b}= -\delta^{\alpha\beta}\delta^c_a +
\epsilon^{\alpha\beta\gamma}J^{\gamma\ c}_{\ a}. 
  \quad\quad\quad \alpha=1,2,3,   \label{Jalgebra}
\end{eqnarray}
and the holonomy group in the tangent space is $Sp(1)\otimes Sp(n)$. 
We define a set of vielbein $1$-forms
\begin{eqnarray}
 d\phi^a e^{\ i\mu}_a, \quad\quad i=1,2, \quad \mu=1,2,\cdots,2n,
\nonumber
\end{eqnarray}
which satisfy 
\begin{eqnarray}
e^{\ i\mu}_{a}e_{\ i\mu}^{b} = \delta^b_a, \quad\quad\quad 
e^{\ i\mu}_{a}e_{\ j\nu}^{a} = \delta^i_j\delta^\mu_\nu.  \nonumber
\end{eqnarray}
The  holonomy groups $Sp(1)$ and $Sp(n)$  act on  the indices $i$ and $\mu$ respectively.  
Using these vielbeins we can construct the triplet complex structure as 
\begin{eqnarray}
\vec J_a^{\ b} = -ie^{\ i\mu}_{a}\vec \sigma_i^{\ j}e_{\ j\mu}^{b}.
 \label{triplet}
\end{eqnarray}
Lower or raise the indices as $\vec J_a^{\ b}g_{bc}\equiv \vec J_{ac}$ and $g^{ab}\vec J_b^{\ c}=\vec J^{ac}$. Then it is easy to show that 
$
\vec J_{ab}=-\vec J_{ba}, \ \vec J^{ab}=-\vec J^{ba}.
$ 
We postulate that the vielbeins are covariantly constant:
\begin{eqnarray}
\nabla_a e^{\ i\mu}_{b}= \partial_a e^{\ i\mu}_{b} -\Gamma^{\ \ c}_{ab}
e^{\ i\mu}_{c} + \omega_{a j\nu}^{ \ \ i\mu}e^{\ j\nu}_{b}  =0.  \label{pos}
\end{eqnarray}
Here $\omega_{a j\nu}^{ \ \ i\mu} $ is the spin connection 
 Lie-valued in $Sp(1)\otimes Sp(n)$, so that 
\begin{eqnarray}
 \omega_{a j\nu}^{ \ \ i\mu} = \omega_{a j}^{\ \ i} \delta_\nu^\mu + \omega_{a \nu}^{\ \ \mu} \delta_j^i.     \label{spin}
\end{eqnarray}
We solve (\ref{pos}) for the spin conections $\omega_{a j}^{\ \ i}$ and $\omega_{a \nu}^{\ \ \mu}$ to find 
\begin{eqnarray}
 \omega_{a j}^{\ \ i} = -{1\over 2n}e^{b}_{\ j\nu}{\mathop{\nabla}^\circ}_a e^{\ i\nu}_{b},      \quad\quad\quad
 \omega_{a \nu}^{\ \ \mu} = - {1\over 2} e^{b}_{\ j\nu}{\mathop{\nabla}^\circ}_a e^{\ j\mu}_{b}.  \nonumber
\end{eqnarray}
Here $\displaystyle{\mathop{\nabla}^\circ}_c$ is the covariant derivative which does not contain the spin connection. 
From the postulate it follows that the metric and the triplet complex structure  are covariantly constant:
\begin{eqnarray}
\nabla_a g_{bc} &=& \partial_a g_{bc}- \Gamma^{\ \ e}_{ab}g_{ec}-\Gamma^{\ \ e}_{ac}g_{be} 
=0,  \label{covg} \\
\nabla_a \vec J_b^{\ c} &=& \partial_a \vec J_b^{\ c} - \Gamma^{\ \ e}_{ab}\vec J^{\ \ c}_{\ e} + \Gamma^{\ \ c}_{ae}\vec J_{b}^{\ e} + \vec\omega_a\times \vec J_b^{\ c} = 0, 
\label{covJ}
\end{eqnarray}
with
$
\vec\omega_a \equiv -i\vec\sigma_i^{\ j}\omega_{a j}^{\ \ i}. 
$
The Riemann curvature tensor is given by 
$$
R_{abc}^{\ \ \ d} = \partial_b\Gamma_{ac}^{\ \ d} -\partial_a\Gamma_{bc}^{\ \ d} 
- \Gamma_{bc}^{\ \ e} \Gamma_{ae}^{\ \ d}+\Gamma_{ac}^{\ \ e} \Gamma_{be}^{\ \ d}.
$$
Correponding to the spin connection (\ref{spin}) it decomposes in the tangent space:
\begin{eqnarray}
R_{abc}^{\ \ \ d}e^{\ i\mu}_{d}e_{\ j\nu}^{c} = R_{abj}^{\ \ \ i} \delta_\nu^\mu + R_{ab\nu}^{\ \ \ \mu} \delta_j^i.   \nonumber
\end{eqnarray}
We may write  the $Sp(1)$ curveture $R_{abj}^{\ \ \ i}$ as
$
\vec R_{ab} = -i\vec\sigma_i^{\ j}R_{abj}^{\ \ \ i},   
$
or equivalently  as 
\begin{eqnarray}
\vec R_{ab} = -{1\over 2n}R_{abc}^{\ \ \ d}\vec J_d^{\ c}={1\over 2n}R_{abcd}\vec J^{cd}
= {1\over n}R_{cab}^{\ \ \ d}\vec J_d^{\ c}.  \label{vecR}
\end{eqnarray}
From the postulate (\ref{pos}) it is given by  
\begin{eqnarray}
\vec R_{ab}=\partial_{[a}\vec \omega_{b]} + \vec \omega_a\times \vec\omega_b.
 \label{kkk}
\end{eqnarray}
In \cite{BCWG} they have shown various relations among $\vec R_{ab}, \vec J_a^{\ b}, R_{ab}(\equiv -R_{cabd}g^{cd})$ and $R(\equiv $  
$g^{ab}R_{ab})$:
\begin{eqnarray}
R_{ab}&=& -{1\over 3}(n+2)\vec J_a^{\ c} \cdot \vec R_{cb}, \label{RicciJ}\\
\vec R_{ab} &=& {1\over n+2}\vec J_a^{\ c}R_{cb} = {1\over 2}\vec J_a^{\ c}\times \vec R_{cb},  \label{veccurva}    \\
R_{ab} &=& {1\over 4n}R g_{ab}. \label{relation}
\end{eqnarray}
Combining (\ref{veccurva}) and (\ref{relation}) gives 
\begin{eqnarray}
 \vec R_{ab} = \nu \vec J_{ab}. \label{vecRicciJ}
\end{eqnarray}

When the quaternionic K\"ahler manifold is a coset space $G/H$, there is a set of the Killing vectors $R^{Aa}=(R^{1a},R^{2a},\cdots,R^{Na})$ with $N= {\rm dim}\ G$ which represents the isometry group $G$. They are required to satisfy 
\begin{eqnarray}
{\cal L}_{R^A} R^{Ba} &\equiv& R^{Ab}\partial_b R^{Ba}- R^{Bb}\partial_b R^{Aa}
 = f^{ABC}R^{Ca},  \label{symmetry}  \\
{\cal L}_{R^A} g_{ab} &\equiv& R^{Ac}\partial_c g_{ab} + \partial_a R^{Ac}g_{cb} + \partial_b R^{Ac}g_{ca} =0,   \label{Lie-g} \\
{\cal L}_{R^A}\vec J_a^{\ b} &\equiv& R^{Ac}\partial_c \vec J_a^{\ b} + \partial_a R^{Ac}\vec J_c^{\ b}- \partial_c R^{Ab}\vec J_a^{\ c}  =  \vec r^A \times \vec J_a^{\ b}, \label{Lie-J}
\end{eqnarray}
in which ${\cal L}_{R^A}$ is the Lie-variation with respect to $R^A$. 
For covariant derivatives of the Killing vectors  we have 
\begin{eqnarray}
\nabla_a R^A_{\ b} + \nabla_b R^A_{\ a} &=& 0,   \label{Killing}   \\
\nabla_a\nabla_b R^A_c &=&  -R_{bca}^{\ \ \ d}R^A_{\ d}, \label{Killing'} \\
\nabla_a\nabla_b R_{\ c}^A +\nabla_b\nabla_c R_{\ a}^A +\nabla_c\nabla_a R_{\ b}^A &=& 0.  \label{cyclicDR} \\
R^A_{\ a}\nabla_b R^A_{\ c} &=& -{1\over 2}f^{ABC}R^A_{\ a}R^B_{\ b}R^C_{\ c},
  \label{RcovR} 
\end{eqnarray}
Eq. (\ref{Killing}) is  the Killing equation which is equivalent  to (\ref{Lie-g}). 
Eq. (\ref{Killing'}) follows by  calculating the {\it l.h.s.} as
\begin{eqnarray}
\nabla_a\nabla_b R^A_{\ c} &=& {1\over 2}\nabla_{\{a}\nabla_{b\}} R^A_{\ c} 
 +{1\over 2}\nabla_{[a}\nabla_{b]} R^A_{\ c}   \nonumber \\
   &=& {1\over 2}(-\nabla_{[a}\nabla_{c]} R^A_{\ b} -\nabla_{[b}\nabla_{c]} R^A_{\ a} )  
 +{1\over 2}\nabla_{[a}\nabla_{b]} R^A_{\ c}   \nonumber \\
  &=& {1\over 2}(-R_{acb}^{\ \ \ d} -R_{bca}^{\ \ \ d} + R_{abc}^{\ \ \ d})R^A_{\ d}, \nonumber
\end{eqnarray}
with   (\ref{Killing}) and  the cyclic property of the Riemann curvature 
tensor 
\begin{eqnarray}
R_{abc}^{\ \ \ d} +R_{bca}^{\ \ \ d} +R_{cab}^{\ \ \ d} = 0. \label{cyclic} 
\end{eqnarray}
Combining (\ref{Killing'}) and (\ref{cyclic}) we obtain (\ref{cyclicDR}). 
 When the coset space $G/H$ is irreducible, it follows from (\ref{Lie-g}) or (\ref{Killing}) that 
\begin{eqnarray}
g_{ab}=R^A_{\ a}R^A_{\ b}. \label{metric}    
\end{eqnarray}
In using this metric formula  our arguments differ from those in the literature. We will  be able to explore  more on the triplet Killing potentials $\vec M^A$.   (See the later arguments for  the reducible case, if there exists any  quaternionic K\"ahler manifold for this case.) 
Put  (\ref{symmetry}) in the covariant form  
\begin{eqnarray}
R^{Ab}\nabla_b R^B_{\ c}- R^{Bb}\nabla_b R^A_{\ c}
 = f^{ABC}R^C_{\ c}.    \label{symmetry'}
\end{eqnarray}
Contract  this equation with $R^A_{\ a}R^B_{\ b}$. Using the formula (\ref{metric}) and its consequence $\nabla_c(R^A_{\ a}$ 
$R^A_{\ b}) = 0$ we then get  (\ref{RcovR}).

Now we proceed to derive (\ref{LieM'})$\sim$(\ref{RiemannCurv'}). We calculate the following the Lie-derivative
\begin{eqnarray}
 {\cal L}_{R^A}\nabla_c R^{Bb} = R^{Aa}\nabla_a \nabla_c R^{Bb} - \nabla_a R^{Ab}\nabla_c R^{Ba} + \nabla_c R^{Aa}\nabla_a R^{Bb}. \label{Lie-dR}
\end{eqnarray}
On the other hand we have 
\begin{eqnarray}
\nabla_c R^{Aa}\nabla_a R^{Bb} + R^{Aa}\nabla_c \nabla_a R^{Bb} - (A\rightleftharpoons B ) = f^{ABC}\nabla_c R^{Bb}.   \nonumber
\end{eqnarray}
from (\ref{symmetry}). 
Eliminating the last two terms in (\ref{Lie-dR}) by this formula and using  (\ref{Killing'}) and (\ref{cyclic}) 
 yields 
\begin{eqnarray}
{\cal L}_{R^A}\nabla_c R^{Bb}&=& R^{Aa}\nabla_{[a} \nabla_{c]} R^{Bb} +
R^{Ba}\nabla_c \nabla_a R^{Ab} + f^{ABC}\nabla_c R^{Bb} \nonumber \\
&=&  f^{ABC}\nabla_c R^{Bb}.  \nonumber
\end{eqnarray}
 With this and (\ref{Lie-J}) the triplet Killing potentials  (\ref{dR}) satisfy the property (\ref{LieM'}). 
Note that 
\begin{eqnarray}
\vec M^A = -{1\over 2n}\vec J^{ab}\nabla_a R^A_{\ b} = -{1\over 2n}\nabla_a(\vec J^{ab} R^A_{\ b})
 = -{1\over 2n\nu}\nabla_b\nabla^b \vec M^A, \label{eigen}
\end{eqnarray}
by (\ref{dM}), (\ref{dR}) and (\ref{covJ}). This implies that $\vec M^A$ is  an eigen vector of the Beltrami-Laplace operator, {\it i.e.} (\ref{BLd'}). 
Note also that 
\begin{eqnarray}
 \nabla_a \vec M^A \cdot \nabla_b \vec M^A = \nu^2\vec J_a^{\ c}R^A_{\ c}\cdot \vec J_b^{\ d}R^A_{\ d},  \label{xxx}
\end{eqnarray}
by (\ref{dM}). Calculate the {\it r.h.s.} with (\ref{Jalgebra}). Then (\ref{xxx}) becomes (\ref{gMM'}) owing to (\ref{metric}). 
Eq. (\ref{ScalarRiemann'}) can be shown by manipulating the formula 
\begin{eqnarray}
\vec J^{ab}\cdot \vec J^{cd}R_{abcd}=-\vec J^{ab}\cdot \vec J^{cd}R^A_{\ d}\nabla_c\nabla_a R^A_{\ b},  \label{www}
\end{eqnarray}
which is obvious from   (\ref{Killing'}). 
By (\ref{dM}) and (\ref{dR})  the {\it r.h.s.}  becomes 
\begin{eqnarray}
-{\nu\over 2n}\vec J^{ab}\cdot\vec J^{cd}R^A_{\ d}\nabla_c\nabla_a R^A_{\ b}
 &=& \nabla^c \vec M^A\cdot \nabla_c \vec M^A    \nonumber
  \\
&=& \nabla^c[\partial_c({1\over 2}\vec M^A\vec M^A)] - \vec M^A \cdot \nabla^c \nabla_c\vec M^A.  \label{ppp}
\end{eqnarray}
From the property (\ref{LieM'}) we have ${\cal L}_{R^B}(\vec M^A\cdot\vec M^A)=0$. Contracting this equation with $R^{Bb}$  and using (\ref{metric}) 
yields $g^{ba}\partial_a (\vec M^A\cdot\vec M^A)=0$ . If the metric is non-degenerate, we get 
$\vec M^A\cdot\vec M^A$ to be a constant.    
Then (\ref{ppp}) becomes 
\begin{eqnarray}
\vec J^{ab}\cdot\vec J^{cd}R^A_{\ d}\nabla_c\nabla_a R^A_{\ b}= -4n^2\vec M^A\cdot \vec M^A,   \label{JJR}
\end{eqnarray}
by (\ref{eigen}). 
On the other hand the {\it l.h.s.} of (\ref{www}) is calculated as 
$$
\vec J^{ab}\cdot\vec J^{cd}R_{abcd} = 2n\vec J^{ab}\cdot \vec R_{ab} = {6n\over n+2}R,
$$
by (\ref{Jalgebra}), (\ref{vecR}) and (\ref{vecRicciJ}). With this and  (\ref {JJR})  the formula (\ref{www})  becomes (\ref{ScalarRiemann'}).
 Finally we show (\ref{RiemannCurv'}). It consists of the two equalities. To show the  equality 
in the first line we rewrite  formula 
\begin{eqnarray}
\nabla_{[a}\nabla_{b]}\vec J_{cd} + \nabla_{[c}\nabla_{d]}\vec J_{ab} = 0,
 \nonumber
\end{eqnarray}
which is obvious from (\ref{covJ}) 
as 
\begin{eqnarray}
R_{abce}\vec J_d^{\ e} + R_{abed}\vec J_c^{\ e} + R_{aecd}\vec J_b^{\ e} +R_{ebcd}\vec J_a^{\ e} = 0,  \label{100}
\end{eqnarray}
by  (\ref{kkk}) and  (\ref{vecRicciJ}). Multiply both sides of (\ref{100}) by $\vec J_f^{\ d}$. Using (\ref{Jalgebra}) and (\ref{Killing'}) we get
\begin{eqnarray}
-3R_{abcf} = R^A_{\ d}\nabla_e\nabla_a R^A_{\ b}\vec J_c^{\ e}\cdot \vec J_f^{\ d}
+ R^A_{\ d}\nabla_c\nabla_a R^A_{\ e}\vec J_b^{\ e}\cdot\vec J_f^{\ d}
+R^A_{\ d}\nabla_c\nabla_e R^A_{\ b}\vec J_a^{\ e}\cdot\vec J_f^{\ d}.
\nonumber
\end{eqnarray}
This becomes 
\begin{eqnarray}
-3R_{abcf} = -R^A_{\ d}\nabla_{[a}\nabla_{b]} R^A_{\ e}\vec J_c^{\ e}\cdot \vec J_f^{\ d}
+ R^A_{\ d}\nabla_c\nabla_a R^A_{\ e}\vec J_b^{\ e}\cdot\vec J_f^{\ d}
-R^A_{\ d}\nabla_c\nabla_b R^A_{\ e}\vec J_a^{\ e}\cdot\vec J_f^{\ d}, 
\nonumber 
\end{eqnarray}
by (\ref{Killing} ) and (\ref{cyclicDR}). Using (\ref{dM}) in the {\it r.h.s.} of the last equation leads us to see  that the  equality in the first line of (\ref{RiemannCurv'}) indeed holds. To show the second equality in (\ref{RiemannCurv'})   we write the Riemann curvature tensor in the form 
\begin{eqnarray}
 R_{abcd}= -\nabla_c(R^A_{\ d}\nabla_a R^A_{\ b}) +(\nabla_c R^A_{\ d})(\nabla_a R^A_{\ b}), \label{alterR}
\end{eqnarray}
by (\ref{Killing'}). Calculate the first term of the {\it r.h.s.} by  (\ref{RcovR}). We then see that $R_{abcd}$  is expressed in terms of $\nabla_b R^A_{\ c}$ alone. Note the formula 
\begin{eqnarray}
\nabla_b R^A_{\ c} = - f^{ABC}R^B_{\ b}R^C_{\ c} + 
{1\over 2}R^{Aa}f_{abc}, \quad\quad f_{abc}\equiv f^{ABC}R^A_{\ a}R^B_{\ b}R^C_{\ c},  \nonumber
\end{eqnarray}
which can be shown  by contracting  (\ref{symmetry'}) with $R^B_{\ b}$ and using (\ref{RcovR}) and (\ref{metric}). By this formula and the Jacobi identity
 for the structure constants $f^{ABC}$, 
the Riemann curvature tensor (\ref{alterR}) gets expressed as given in (\ref{RiemannCurv'}).

We illustrate our general arguments for  quaternionic K\"ahler group manifolds, taking the coset space $Sp(n+1)/Sp(n)\otimes Sp(1)$\cite{AH} as an example. The isometry group $Sp(n+1)$ is generated by operators $X^{IJ}(=X^{JI}), I,J =1,2,\cdots,2(n+1)$, satisfying the Lie-algebra
\begin{eqnarray}
[X^{IJ},X^{KL}]=\epsilon^{IK}X^{JL}+\epsilon^{JL}X^{IK}+\epsilon^{IL}X^{JK}
 + \epsilon^{JK}X^{IL}.    \nonumber
\end{eqnarray}
Here $\epsilon^{IJ}$ is a constant anti-symmetric tensor, which we take to be 

\newfont{\bg}{cmr10 scaled \magstep4}
\newcommand{\bigzerol}{\smash{\hbox{\bg 0}}}
\newcommand{\bigzerou}{%
  \smash{\lower1.7ex\hbox{\bg 0}}}

\begin{eqnarray}
\epsilon^{IJ} = -\epsilon^{JI}= -\epsilon_{IJ}=\epsilon_{JI}=
 \left(
\begin{array}{ccccc}
0 & 1 &  &  & \bigzerou   \\ 
 -1 & 0 &  &  &   \\  
    &  & \ddots &  &  \\
    &  &  & 0 & 1 \\
 \bigzerol  & & & -1& 0 
\end{array}\right),      \nonumber
\end{eqnarray}
for convenience. The generators are decomposed as
$$
\{ X^{IJ}\} = \{ X^{i\mu},X^{\mu\nu},X^{ij}\}, \quad\quad \mu,\nu =1,2,\cdots,2n, \quad\quad    i,j=1,2,
$$
in which $X^{\mu\nu}$ and $X^{ij}$  are generators of the homogeneous group $Sp(n+1)\otimes Sp(1)$ and $X^{i\mu}(= X^{\mu i})$ are broken generators. The quadratic Casimir takes the form 
\begin{eqnarray}
{1\over 2}X^{IJ}X^{KL}\epsilon_{IK}\epsilon_{JL}
= X^{i\mu}X^{j\nu}\epsilon_{ij}\epsilon_{\mu\nu}
 + {1\over 2}X^{\mu\nu}X^{\rho\sigma}\epsilon_{\mu\rho}\epsilon_{\nu\sigma}
 +{1\over 2}X^{ij}X^{kl}\epsilon_{ik}\epsilon_{jl}. \nonumber
\end{eqnarray}
The coset space $Sp(n+1)/Sp(n)\otimes Sp(1)$ is locally parametrized by coordinates $\mbox{$\phi$}^{\mbox{\scriptsize\boldmath{${i\mu}$}}}$ corresponding to the broken generators $X^{i\mu}$. The line element is given by
$$
{\mbox{$ds^2$}}\ \mbox{$=$}\ {\mbox{$g$}}_{{\mbox{\scriptsize{\boldmath $i\mu$},{\boldmath $j\nu$}}}}
\mbox{$d$}{\mbox{$\phi$}}^{{\mbox{\scriptsize{\boldmath $i\mu$}}}}
\mbox{$d$}{\mbox{$\phi$}}^{{\mbox{\scriptsize{\boldmath $j\nu$}}}}.
$$  
We  find the Killing vectors as non-linear realization of 
the Lie-algebra (\ref{symmetry}):
\begin{eqnarray}
\mbox{$R$}^{\mbox{\scriptsize{$(k\sigma)$\boldmath $i\mu$}}}
 &\equiv&
\mbox{$[X$}^{\mbox{\scriptsize{$k\sigma$}}},{\mbox{$\phi$}}^{\mbox{\scriptsize{\boldmath $i\mu$ }}}\mbox{$]$}\ \mbox{$=$}\ \mbox{$\epsilon$}^{\mbox{\scriptsize{$k$\boldmath $i$}}}\mbox{$\epsilon$}^{\mbox{\scriptsize{$\sigma$\boldmath$\mu$}}} \mbox{$+$} \mbox{$\phi$}^{\mbox{\scriptsize{$k$\boldmath $\mu$}}}\mbox{$\phi$}^{\mbox{\scriptsize{{\boldmath $i$}$\sigma$ }}}\mbox{$,$}  \nonumber \\
\mbox{$R$}^{\mbox{\scriptsize{$(\rho\sigma)$\boldmath $i\mu$}}}
 &\equiv&
\mbox{$[X$}^{\mbox{\scriptsize{$\rho\sigma$}}},{\mbox{$\phi$}}^{\mbox{\scriptsize{\boldmath $i\mu$ }}}\mbox{$]$}\ \mbox{$=$}\ \mbox{$\epsilon$}^{\mbox{\scriptsize{$\rho$\boldmath $\mu$}}}\mbox{$\phi$}^{\mbox{\scriptsize{{\boldmath$i$}$\sigma$}}} \mbox{$+$}
\mbox{$\epsilon$}^{\mbox{\scriptsize{$\sigma$\boldmath $\mu$}}}\mbox{$\phi$}^{\mbox{\scriptsize{{\boldmath$i$}$\rho$}}}\mbox{$,$}    \label{Spkilling} \\
\mbox{$R$}^{\mbox{\scriptsize{$(kl)$\boldmath $i\mu$}}}
 &\equiv&
\mbox{$[X$}^{\mbox{\scriptsize{$kl$}}}\ ,{\mbox{$\phi$}}^{\mbox{\scriptsize{\boldmath $i\mu$ }}}\mbox{$]$}\ \mbox{$=$}\ \mbox{$\epsilon$}^{\mbox{\scriptsize{$k$\boldmath $i$}}}\mbox{$\phi$}^{\mbox{\scriptsize{$l${\boldmath$\mu$}}}} \mbox{$+$}\ 
\mbox{$\epsilon$}^{\mbox{\scriptsize{$l$\boldmath $i$}}}\mbox{$\phi$}^{\mbox{\scriptsize{$k${\boldmath$\mu$}}}}\mbox{$.$}   \nonumber
\end{eqnarray}
Then the metric is obtained from (\ref{metric})
\begin{eqnarray}
{\mbox{$g$}}^{{\mbox{\scriptsize{\boldmath $i\mu$},{\boldmath $j\nu$}}}}
&=& \mbox{$[\epsilon(1-\epsilon\phi\epsilon\phi)]$}^{\mbox{\scriptsize{\boldmath$ij$}}}\mbox{$[\epsilon(1-\epsilon\phi\epsilon\phi)]$}^{\mbox{\scriptsize{\boldmath$\mu\nu$}}}\mbox{$,$} \nonumber \\
{\mbox{$g$}}_{{\mbox{\scriptsize{\boldmath $i\mu$},{\boldmath $j\nu$}}}}
&=& \mbox{$[(1-\epsilon\phi\epsilon\phi)^{-1}\epsilon]$}_{\mbox{\scriptsize{\boldmath$ij$}}}\mbox{$[(1-\epsilon\phi\epsilon\phi)^{-1}\epsilon]$}_{\mbox{\scriptsize{\boldmath$\mu\nu$}}}\mbox{$.$} \nonumber
\end{eqnarray}
Here one should undestand matrix multiplication such that 
\begin{eqnarray}
\mbox{$(\epsilon\phi)$}_{\mbox{\scriptsize{\boldmath$i$}}}^{\mbox{\scriptsize{\ \boldmath$\mu$}}}\mbox{$= \epsilon$}_{\mbox{\scriptsize{\boldmath$ik$}}} \mbox{$\phi$}^{\mbox{\scriptsize{\boldmath$k\mu$}}}\mbox{$,$} 
&\quad&\quad 
\mbox{$(\epsilon\phi)$}_{\mbox{\scriptsize{\boldmath$\mu$}}}^{\mbox{\scriptsize{\ \ \boldmath$i$}}}
\mbox{$=\epsilon$}_{\mbox{\scriptsize{\boldmath$\mu\rho$}}}
\mbox{$\phi$}^{\mbox{\scriptsize{\boldmath$i\mu$}}}\mbox{$,$} \nonumber \\  
\mbox{$(\epsilon\phi\epsilon\phi)$}_{\mbox{\scriptsize{\boldmath$i$}}}^{\mbox{\scriptsize{\ \boldmath$j$}}} \mbox{$= \epsilon$}_{\mbox{\scriptsize{\boldmath$ik$}}}\mbox{$\phi$}^{\mbox{\scriptsize{\boldmath$k\rho$}}}\mbox{$\epsilon$}_{\mbox{\scriptsize{\boldmath$\rho\sigma$}}}\mbox{$\phi$}^{\mbox{\scriptsize{\boldmath$j\sigma$}}}\mbox{$,$}
&\quad& \quad  
\mbox{$(\epsilon\phi\epsilon\phi)$}_{\mbox{\scriptsize{\boldmath$\mu$}}}^{\mbox{\scriptsize{\ \boldmath$\nu$}}} 
\mbox{$= \epsilon$}_{\mbox{\scriptsize{\boldmath$\mu\rho$}}}
\mbox{$\phi$}^{\mbox{\scriptsize{\boldmath$k\rho$}}}
\mbox{$\epsilon$}_{\mbox{\scriptsize{\boldmath$kl$}}}
\mbox{$\phi$}^{\mbox{\scriptsize{\boldmath$l\nu$}}}\mbox{$.$}  \nonumber 
\end{eqnarray}
We calculate the Affine connection as
\begin{eqnarray}
\mbox{$\Gamma$}_{{\mbox{\scriptsize{\boldmath $i\mu$},{\boldmath $j\nu$}}}}^{\mbox{\scriptsize{\ \ \boldmath$k\lambda$}}}\mbox{$=\delta$}_{\mbox{\scriptsize{\boldmath$j$}}}^{\mbox{\scriptsize{\boldmath$k$}}}\mbox{$\delta$}_{\mbox{\scriptsize{\boldmath$\mu$}}}^{\mbox{\scriptsize{\boldmath$\lambda$}}}
\mbox{$[(1-\epsilon\phi\epsilon\phi)\epsilon\phi\epsilon]$}_{\mbox{\scriptsize{\boldmath$i\nu$}}}
\mbox{$+\delta$}_{\mbox{\scriptsize{\boldmath$i$}}}^{\mbox{\scriptsize{\boldmath$k$}}}\mbox{$\delta$}_{\mbox{\scriptsize{\boldmath$\nu$}}}^{\mbox{\scriptsize{\boldmath$\lambda$}}}
\mbox{$[(1-\epsilon\phi\epsilon\phi)\epsilon\phi\epsilon]$}_{\mbox{\scriptsize{\boldmath$j\mu$}}}\mbox{$.$}     \label{SpGamma}
\end{eqnarray}
The vielbeins are given by 
\begin{eqnarray}
\mbox{$e$}^{\ \mbox{\scriptsize{$i\mu$}}}
_{\mbox{\scriptsize{\boldmath$i\mu$}}}
\mbox{$= [(1-\epsilon\phi\epsilon\phi)^{-{1\over 2}}]$}_{\mbox{\scriptsize{\boldmath$i$}}}^{\ \mbox{\scriptsize{$i$}}}\mbox{$[(1-\epsilon\phi\epsilon\phi)^{-{1\over 2}}]$}_{\mbox{\scriptsize{\boldmath$\mu$}}}^{\ \mbox{\scriptsize{$\mu$}}}
\mbox{$,$}        \nonumber
\end{eqnarray}
so as to satisfy 
$$
{\mbox{$g$}}_{{\mbox{\scriptsize{\boldmath $i\mu$},{\boldmath $j\nu$}}}}
\ \mbox{$=$}\mbox{$e$}^{\ \mbox{\scriptsize{$i\mu$}}}
_{\mbox{\scriptsize{\boldmath$i\mu$}}}
\mbox{$e$}^{\ \mbox{\scriptsize{$j\nu$}}}
_{\mbox{\scriptsize{\boldmath$j\nu$}}}\mbox{$\epsilon$}_{\mbox{\scriptsize{$ij$}}}\mbox{$\epsilon$}_{\mbox{\scriptsize{$\mu\nu$}}}\mbox{$.$}
$$
Using the formula
$$
\mbox{$(\phi\epsilon\phi)$}^{\mbox{\scriptsize{\boldmath$ij$}}}
\mbox{$=\phi$}^{\mbox{\scriptsize{\boldmath$i\rho$}}}
\mbox{$\epsilon$}_{\mbox{\scriptsize{\boldmath$\rho\sigma$}}}
\mbox{$\phi$}^{\mbox{\scriptsize{\boldmath$j\sigma$}}}\mbox{$={1\over 2}\epsilon$}^{\mbox{\scriptsize{\boldmath$ij$}}}\mbox{$\phi^2$}\mbox{$,$}
$$
with $\mbox{$\phi^2=\epsilon$}_{\mbox{\scriptsize{\boldmath$ji$}}}\mbox{$\phi$}^{\mbox{\scriptsize{\boldmath$i\rho$}}}
\mbox{$\epsilon$}_{\mbox{\scriptsize{\boldmath$\rho\sigma$}}}
\mbox{$\phi$}^{\mbox{\scriptsize{\boldmath$j\sigma$}}}$, we rewrite the vielbeins in the form 
$$
\mbox{$e$}^{\ \mbox{\scriptsize{$i\mu$}}}
_{\mbox{\scriptsize{\boldmath$i\mu$}}}
\mbox{$=\delta$}_{\mbox{\scriptsize{\boldmath$i$}}}^{\mbox{\scriptsize{$i$}}}
\mbox{$[(1-\epsilon\phi\epsilon\phi)]$}_{\mbox{\scriptsize{\boldmath$\mu$}}}^{\ \mbox{\scriptsize{$\mu$}}}
\mbox{$(1-{1\over 2}\phi^2)^{-{1\over 2}}$}\mbox{$.$} 
$$
Then the triplet complex structure (\ref{triplet}) becomes 
\begin{eqnarray}
\mbox{$\vec J$}_{\mbox{\scriptsize{\boldmath$i\mu$}}}^{\mbox{\scriptsize{\ \ \boldmath$j\nu$}}} \mbox{$= -i\vec \sigma$}_{\mbox{\scriptsize{\boldmath$i$}}}^{\mbox{\scriptsize{\ \boldmath$j$}}}\mbox{$\delta$}_{\mbox{\scriptsize{\boldmath$\mu$}}}^{\mbox{\scriptsize{\boldmath$\nu$}}}\mbox{$.$}    \label{Sptriplet}
\end{eqnarray}
Using (\ref{dR}) with (\ref{Spkilling}), (\ref{SpGamma}) and (\ref{Sptriplet}) we calculate the triplet Killing potentials to find
\begin{eqnarray}
\vec M^{k\sigma}=- 2i{(\phi\vec\sigma)^{\rho k}\over 1-{1\over 2}\phi^2}, \quad\quad 
\vec M^{\rho\sigma}= 2i{(\phi\vec\sigma\epsilon\phi)^{\rho\sigma}\over 1-{1\over 2}\phi^2},   \quad\quad 
\vec M^{kl}=-2i{(\epsilon\vec\sigma)^{kl}\over 1-{1\over 2}\phi^2}. \nonumber 
\end{eqnarray}
The Casimir product of $\vec M^A$ is indeed constant:
$$
\vec M^{IJ}\vec M^{KL} \epsilon_{IK}\epsilon_{JL} = 12.
$$
It is easy to check that the triplet Killing potentials have the property (\ref{LieM'}) with
$$
\vec r^{k\sigma}= -i(\phi\vec \sigma)^{k\sigma},\quad\quad   
\vec r^{\rho\sigma}= 0, \quad\quad   
 \vec r^{kl}= -2i (\epsilon\vec \sigma)^{kl}.
$$
On the other hand these $\vec r$'s may be obtained  from (\ref{Lie-J}) as well. Of course they coincide with each other.

The derivations of (\ref{gMM'})$\sim$(\ref{RiemannCurv'}) relied on the metric formula (\ref{metric}). This formula holds at least for a class of quaternionic K\"ahler  coset spaces, called the Wolf space\cite{W}, for which $Sp(n+1)/Sp(n)\otimes Sp(1)$ is the simplest example. To verify this  we use the CCWZ formalism\cite{CCWZ}. The Wolf space is a compact coset space of type $G/S\otimes Sp(1)$ or $G/S\otimes SU(2)$ with  particularly chosen  $G$ and $S$. The generators of $G$, denoted by $T^A$, are decomposed as 
\begin{eqnarray}
\{ T^A \} &=& \{ X^{i\mu}, S^I, Q^\alpha\},    \nonumber \\ 
 i=1,2, \ \ \  \mu=1,2,&\cdots& 2n, \ \ \  I=1,2,\cdots, {\rm dim}\ K,    
\ \ \  \alpha=1,2,3. \nonumber
\end{eqnarray}
Here $S^I$  are generators of the homogeneous group $S$, while $Q^\alpha$ those of $Sp(1)$ or $SU(2)$ depending on the type.  $X^{i\mu} $  are broken generators and transform as
 a tensor in the representation $(\underline 2,\underline {2n})$ under the homogeneous group $Sp(1)\otimes S$ or $SU(2)\otimes S$. In general it is said that the coset space is irreducible(reducible) when the representation of $X^{i\mu}$ is  irreducible(reducible) under $S$. The Wolf space is irreducible. 
The quadratic Casimir of $G$ is given by 
\begin{eqnarray}
T^A T^A = X^{i\mu}X^{j\nu}\eta_{i\mu,j\nu} + S^IS^I +Q^\alpha Q^\alpha,
 \nonumber
\end{eqnarray}
in which  $\eta_{i\mu,j\nu}$ is the Killing form for the coset  part, {\it i.e.}, $\eta_{i\mu,j\nu} = \epsilon_{ij}\epsilon_{\mu\nu}$ for the previous example $Sp(n+1)/Sp(n)\otimes Sp(1)$. 
The coset space is locally para\-metrized by $4n$ real coordinates 
$\mbox{$\phi$}^{\mbox{\scriptsize\boldmath{${i\mu}$}}}$ 
 corresponding to the broken generators $X^{i\mu}$. Cosider the quantity $\mbox{$U$}\ \mbox{$=$}\ \mbox{$\exp$}\mbox{$(\phi$}^{\mbox{\scriptsize{\boldmath $i\mu$}}}\mbox{$X$}^{j\nu}\mbox{$\eta$}_{\mbox{\scriptsize{{\boldmath $i$$\mu$},$j\nu$}}}\mbox{$)$}$. By left multiplication of $\exp(\epsilon^A T^A)\in  G$ we find 
\begin{eqnarray}
\exp(\epsilon^A T^A)U(\phi)= U(\phi')h(\phi,g),  \label{trans}
\end{eqnarray}
appropriately choosing the conpensator $h(\phi,g)\in Sp(1)\otimes S$ or $SU(2)\otimes S$. Here $\epsilon^A$ are global parameters. When they are infinitesimal (\ref{trans}) yields the Killing vectors $\mbox{$R$}^{\mbox{\scriptsize{$A$\boldmath$i\mu$}}}\mbox{$(\phi)$}$ as
$$
\mbox{$\delta \phi$}^{\mbox{\scriptsize{\boldmath $i\mu$}}}\ \mbox{$=$}\ 
\mbox{$\phi'$}^{\mbox{\scriptsize{\boldmath $i\mu$}}}
-\mbox{$\phi$}^{\mbox{\scriptsize{\boldmath $i\mu$}}}
\ \mbox{$=$}\  \mbox{$\epsilon^A$}
\mbox{$R$}^{\mbox{\scriptsize{$A$\boldmath$i\mu$}}}\mbox{$(\phi)$}\mbox{$.$}
$$
The fundamental object to construct the metric is the Cartan-Maurer $1$-form $ U^{-1}dU $. It is valued in the Lie-algebra of $G$ as
$$
 U^{-1}dU= e^{i\mu}X^{i\mu}\eta_{i\mu,j\nu} + \omega^I S^I + \omega^\alpha Q^\alpha. 
$$  
 $e^{i\mu}$ is a vielbein $1$-form, while $\omega^I$ and $\omega^\alpha$ connection $1$-forms corresponding to the respective holonomy groups $S$ and $Sp(1)$. With this vielbein $1$-form the metric ${\mbox{$g$}}_{{\mbox{\scriptsize{\boldmath $i\mu$},{\boldmath $j\nu$}}}}$ is given 
by 
\begin{eqnarray}
{\mbox{$g$}}_{{\mbox{\scriptsize{\boldmath $i\mu$},{\boldmath $j\nu$}}}}
\ \mbox{$=$}\ \mbox{$e$}^{\ \mbox{\scriptsize{$i\mu$}}}
_{\mbox{\scriptsize{\boldmath$i\mu$}}}
\mbox{$e$}^{\ \mbox{\scriptsize{$j\nu$}}}
_{\mbox{\scriptsize{\boldmath$j\nu$}}}
\mbox{$\eta$}_{\mbox{\scriptsize{$i\mu,j\nu$}}}\mbox{$.$} 
\label{metricCCWZ}
\end{eqnarray}
It is also given by (\ref{metric}) {\it i.e.}, ${\mbox{$g$}}_{{\mbox{\scriptsize{\boldmath $i\mu$},{\boldmath $j\nu$}}}}\ \mbox{$=$}\ 
\mbox{$R$}^{\mbox{\scriptsize{$A$}}}_{\ \mbox{\scriptsize{\boldmath$i\mu$}}}
\mbox{$R$}^{\mbox{\scriptsize{$A$}}}_{\ \mbox{\scriptsize{\boldmath$j\nu$}}}$.  Both metrics are equivalent because the value at the origin $\mbox{$g$}_{{\mbox{\scriptsize{\boldmath $i\mu$},{\boldmath $j\nu$}}}}\mbox{$|_{\phi=0}$}$  and the Lie-variation with respect to the Killing vectors $\mbox{${\cal L}_{R^A}$}\mbox{$g$}_{{\mbox{\scriptsize{\boldmath $i\mu$},{\boldmath $j\nu$}}}}$ are the same.  Equivalently we can say that
\begin{eqnarray}
\mbox{$e$}^{\ \mbox{\scriptsize{$i\mu$}}}
_{\mbox{\scriptsize{\boldmath$i\mu$}}}
\ \mbox{$=$} \
\mbox{$R$}^{\mbox{\scriptsize{$A$}}}_{\ \mbox{\scriptsize{\boldmath$i\mu$}}}
\mbox{$U$}^{\mbox{\scriptsize{$A,i\mu$}}}\mbox{$,$} \label{vielbeinCCWZ}  
\end{eqnarray}
in which $U^{A,i\mu}$ are matrix elements of $U$ in the adjoint representation of $G$\cite{A2}. The point is that the Wolf space is irreducible. For reducible coset spaces  the metric formula (\ref{metricCCWZ}) should be generalized as
\begin{eqnarray}
{\mbox{$g$}}_{{\mbox{\scriptsize{\boldmath $i\mu$},{\boldmath $j\nu$}}}}
\ \mbox{$=$}\  \mbox{$\sum_{\mu=1}^N c_\mu$}
\mbox{$e$}^{\ \mbox{\scriptsize{$i\mu$}}}
_{\mbox{\scriptsize{\boldmath$i\mu$}}}
\mbox{$e$}^{\ \mbox{\scriptsize{$j\nu$}}}
_{\mbox{\scriptsize{\boldmath$j\nu$}}}\mbox{$\eta_{i\mu,j\nu}$}\mbox{$,$}
 \label{genemetric}
\end{eqnarray}
in which $c_\mu$ may take different values for each irreducible component of $X^{i\mu}$ such as $c_\mu = (c_1,\cdots,c_1,c_2,\cdots,c_2,c_3\cdots\cdots,c_N,\cdots,c_N)$ with $\sum_{\mu=1}^N 1 =2n$\cite{A2,KIK}. Accordingly the formula (\ref{metric}) is generalized to the one obtained   by putting  the vielbeins  (\ref{vielbeinCCWZ}) in (\ref{genemetric})\cite{A2}. 
Presumably quaternionic K\"ahler manifolds could exist also 
in such reducible cases similarly to the ordinary K\"ahler manifold\cite{A2,KIK}.

In this letter   quaternionic K\"ahler manifolds were studied in view of an explicit construction of the metric. For the triplet Killing potentials we have shown  the properties (\ref{LieM'})$\sim$(\ref{RiemannCurv'})
 which have been  missing in the literature. Among them the properties (\ref{gMM'})$\sim$(\ref{RiemannCurv'}) were derived with recourse to 
  the metric formula (\ref{metric}). 
 It seems that those properties  have been overlooked  in generality of the arguments with no specification of the  metric. The metric can not be determined  by the Killing equation (\ref{Killing}) alone. One needs the initial condition at the origin of the manifold. By taking account of it 
the metric formula (\ref{metric}) was justified for the irreducible coset space. It is worth studying also the reducible case in the constructive approach of this letter.

\vspace{1cm}
\noindent
\noindent
{\Large\bf Acknowledgements}

I thank R. D'Auria for the hospitality at Politecnico di Torino. I also thank him and A. Van Proeyen for the discussions during the stay through which this work was started. 
The work was supported in part  by the Grant-in-Aid for Scientific Research No.
13135212.

\end{document}